\documentstyle[prd,aps]{revtex}
\begin{document}
\title{
\hfill{\normalsize{\bf TRI-PP-97-20}}\\
\hfill{\normalsize{\bf ZU-TH 18/97}}\\
\hfill{\normalsize{\bf MKPH-T-97-21}}\\[2cm]
Compton Scattering and the Spin Structure of the Nucleon at Low Energies
}
\author{Thomas R. Hemmert$^a$, Barry R. Holstein$^b$, Joachim
Kambor$^c$, and Germar Kn\"{o}chlein$^d$}
\address{$^a$ Theory Division, TRIUMF, 4004 Wesbrook Mall, Vancouver,
BC, Canada V6T 2A3}
\address{$^b$ Department of Physics and Astronomy, University of
Massachusetts, Amherst, MA, USA 01003}
\address{$^c$ Theoretische Physik, Universit\"{a}t Z\"{u}rich, CH-8057
Z\"{u}rich, Switzerland}
\address{$^d$ Institut f\"{u}r Kernphysik, Johannes Gutenberg-Universit\"{a}t,
D-55099 Mainz, Germany}
\maketitle
\begin{abstract}
We analyze polarized Compton scattering which provides information on the
spin-structure of the nucleon. For scattering processes with photon energies
up to 100 MeV the spin-structure dependence can be
encoded into four independent parameters---the so-called spin-polarizabilities
$\gamma_i \; , \; i=1...4$ of the nucleon,
which we calculate within the framework of the ``small scale expansion''
in SU(2) baryon chiral perturbation theory.  Specific application is made to
``forward'' and ``backward'' spin-polarizabilities.
\end{abstract}
\newpage
\narrowtext
\twocolumn

\section{Introduction}
The subject of Compton scattering from the nucleon
has been an active one of late, with
much activity on both experimental and theoretical fronts. Historically,
Compton scattering off a (point) spin 1/2 target with an anomalous magnetic
moment $\kappa$ has been calculated by Powell \cite{Powell} and agrees
reasonably well with experimental cross sections up to photon energies of
50MeV for unpolarized scattering. If one increases the energy of the incoming
photon beam the simple Powell model for a point nucleon fails, as one is
picking up sensitivity to the
internal structure of the nucleon. One can account for this
nucleon structure-dependent effect in {\em unpolarized} Compton
scattering by introducing {\em two} free parameters---commonly denoted
{\em electric} ($\alpha_E$) and {\em magnetic} ($\beta_M$)  polarizabilities.
of the nucleon. It is known that this approach works very well up to photon
energies of 100 MeV or so.

Experimentally these polarizabilities for both neutron
and proton have been recently measured\footnote{The quoted
neutron numbers were
obtained from an analysis of neutron transmission experiments on a Pb
target.  However, a
recent paper quotes a quite different number.\cite{77}}
\begin{eqnarray}
\alpha_E^{(p)}&=&(12.1\pm 0.8\pm 0.5)\times 10^{-4}\,{\rm fm}^3\cite{1}
\label{eq:za}\\
\beta_M^{(p)}&=&(2.1\mp 0.8\mp 0.5)\times 10^{-4}\,{\rm fm}^3\cite{1}\\
\alpha_E^{(n)}&=&(12.6\pm 1.5\pm 2.0)\times 10^{-4}\,{\rm fm}^3\cite{2}\\
\beta_M^{(n)}&=&(3.2\mp 1.5\mp 2.0)\times 10^{-4}\,{\rm fm}^3\cite{2}\label{eq:zd}
\end{eqnarray}
and have been confronted with various theoretical estimates.  For
example, using SU(2) ``heavy baryon'' ChPT to ${\cal O}(p^4)$ a
calculation by Bernard, Kaiser, Schmidt and Mei\ss ner yielded \cite{3}
\begin{eqnarray}
\alpha_E^{(p)}&=&(10.5\pm2.0)\times 10^{-4}\,{\rm fm}^3\label{eq:tha}\\
\beta_M^{(p)}&=&(3.5\pm 3.6)\times 10^{-4}\,{\rm fm}^3\\
\alpha_E^{(n)}&=&(13.4\pm 1.5)\times 10^{-4}\,{\rm fm}^3\\
\beta_M^{(n)}&=&(7.8\pm 3.6)\times 10^{-4}\,{\rm fm}^3 \label{eq:thd}
\end{eqnarray}
in reasonable agreement with the experimental results of
Eqs.(\ref{eq:za}-\ref{eq:zd}).
However, there exist significant theoretical error bars in
Eqs.(\ref{eq:tha}-\ref{eq:thd}), associated with uncertainties
in the estimation of various counterterm contributions via a ``resonance
saturation'' hypothesis. While the concept of
``resonance saturation'' is very well established for the ${\cal
O}(p^4)$ counterterms in the {\em meson} lagrangian \cite{vdm}, its
analogue in the {\em baryon} sector is far more complex due to the
rich structure both in the baryon resonance spectrum and in the
variety of couplings between baryons and meson resonances \cite{aspects}.  
In fact, the largest
uncertainty in Eqs.(\ref{eq:tha}-\ref{eq:thd}) is due to the lowest
lying nucleon resonance---$\Delta$(1232).
In an attempt to understand this contribution
more fully Hemmert, Holstein and Kambor have developed a systematic
``small scale expansion'' \cite{4} within the SU(2) heavy-mass formulation
of baryon ChPT,
which allows treatment of both $\Delta(1232)$ and the nucleon as explicit
degrees of freedom\footnote{The idea of treating the spin 3/2 baryon
resonances as explicit degrees of freedom in ``heavy baryon'' ChPT has
been advocated by Jenkins and Manohar \cite{JM}. For estimates of the
spin-independent polarizabilities using this SU(3) approach see ref.\cite{BSS}
and references therein.}
rather than simply as a static contribution to counterterms.
In this approach, one sets up the calculation as an expansion in the
(small) quantity $\epsilon$, which
collectively denotes non-relativistic momenta $p$, the pion mass
$m_\pi$ or the nucleon-delta mass splitting $\Delta = M_\Delta-M_N$.

Nucleon Compton scattering has been calculated recently
within this ``small scale expansion'' framework to ${\cal O}(\epsilon^3)$
in ref.\cite{5}. However, this calculation suffered from the lack of accurate
information on the strength of the $\pi N\Delta$ and $\gamma N\Delta$ couplings
in the ``small scale expansion'' formalism. Instead, estimates originating
from relativistic Born analyses were used for the couplings. This inadequacy
can now be overcome. Utilizing a new determination \cite{13} of these couplings
within the framework of the ``small scale expansion'' with the results of 
ref.\cite{5} one finds
\begin{eqnarray}
\alpha_E^{(p)}=\alpha_E^{(n)}&=&[12.2({\rm N\pi-loop})+0(\Delta-{\rm
pole})\nonumber \\
& &\phantom{[12.2}
+4.2(\Delta\pi-{\rm loop})]\times 10^{-4}\,{\rm fm}^3 \; , \\
\beta_M^{(p)}=\beta_M^{(n)}&=&[1.2({\rm N\pi-loop})+7.2(\Delta-{\rm
pole})\nonumber \\
& &\phantom{[1.2}
+0.7(\Delta\pi-{\rm loop})]\times 10^{-4}\,{\rm fm}^3 \; .
\end{eqnarray}
The results are still larger than the currently known experimental information
but now much closer to the domain of uncertainty of the chiral $O(p^4)$ 
calculation. Obviously 
$O(\epsilon^4)$ calculations of $\alpha_E, \beta_M$ are called 
for in the ``small scale expansion'' to study the convergence of the 
perturbation series\footnote{The satisfactory agreement
between the
${\cal O}(p^4)$ calculation in ``standard ChPT'' \cite{3} and
the latest experimental results Eqs.(\ref{eq:za}-\ref{eq:zd}) depends
upon a cancelation between counterterms dominated by
$\Delta$(1232) pole terms and higher order $\pi -N$ continuum contributions.
A similar mechanism can presumably be implemented in the ``small scale 
expansion''
approach if one pushes the calculation beyond ${\cal O}(\epsilon^3)$.
A less plausible but 
alternative mechanism for the reduction of the large delta pole contribution
in $\beta_M$ has been proposed in ref.\cite{BSS}.}
and are underway.

Our goal in this note, however, is to extend the discussion on {\em
unpolarized} Compton scattering given so far
to the more general case of {\em polarized} Compton scattering.
In analogy to the unpolarized case one can parameterize the
spin-dependent nucleon structure beyond the anomalous magnetic moment in
terms of unknown parameters $\gamma_i \; , i=1...4$ , which are called the
{\em spin-polarizabilities} of the nucleon.
\cite{Ragusa}.
In section 2 we give a definition for spin-polarizabilities in terms of a
low-energy expansion of the Compton amplitude and in section 3 we present the
ChPT predictions for the $\gamma_i$ to ${\cal O}(\epsilon^3)$ in the ``small
scale expansion'' scheme of ref.\cite{4}. In particular, we have calculated
all contributions to the spin-dependent structure of the nucleon
which arise from $\pi -N$ continuum states, $\Delta$(1232)
pole graphs, $\pi -\Delta$ continuum states and $\pi^0\gamma\gamma$ anomaly
effects to ${\cal O}(\epsilon^3)$. We analyze the underlying physics behind
each $\gamma_i$ and conclude with a discussion of the case of the
so called {\em forward} and {\em backward} spin-polarizabilities of the
nucleon, for which some experimental information is available from multipole
analyses and fits to the {\em unpolarized} cross sections \cite{SWK94,6}.

\section{Spin Dependent Compton Scattering}

Assuming invariance under parity, charge conjugation and time reversal
symmetry the general amplitude for Compton scattering can be written in terms
of six structure dependent functions $A_i(\omega , \theta ), \; i=1..6$,
with $\omega = \omega^\prime$ denoting the photon energy in the c.m. frame and
$\theta$ being the c.m. scattering angle:
\begin{eqnarray}
T &=& A_1(\omega,\theta)\vec{\epsilon}^{* \prime}\cdot\vec{\epsilon}
+A_2(\omega,\theta)\vec{\epsilon}^{* \prime}\cdot\hat{k} \; \vec{\epsilon}
\cdot\hat{k}^\prime \nonumber\\
&+&iA_3(\omega,\theta)\vec{\sigma}\cdot(\vec{\epsilon}^{* \prime}\times
\vec{\epsilon})
+iA_4(\omega,\theta)\vec{\sigma}\cdot(\hat{k}^\prime \times\hat{k})
\vec{\epsilon}^{* \prime} \cdot\vec{\epsilon} \nonumber\\
&+& iA_5(\omega,\theta)\vec{\sigma}\cdot[(\vec{\epsilon}^{* \prime} \times
\hat{k}) \vec{\epsilon}\cdot\hat{k}^\prime -(\vec{\epsilon}\times
\hat{k}^\prime ) \vec{\epsilon}^{* \prime} \cdot\hat{k}]\nonumber\\
&+& iA_6(\omega,\theta)\vec{\sigma}\cdot[(\vec{\epsilon}^{* \prime}\times
\hat{k}^\prime ) \hat{\epsilon}\cdot\hat{k}^\prime -(\vec{\epsilon}\times
\hat{k})\vec{\epsilon}^{* \prime} \cdot\hat{k}]
\end{eqnarray}
Here $\vec{\epsilon},\hat{k}\; (\vec{\epsilon}^\prime ,\hat{k}^\prime )$ are
the polarization vector, direction  of the incident (final) photon while
$\vec{\sigma}$ represents the (spin) polarization vector of the nucleon.

Following general conventions
we separate the pion-pole (``anomalous'')
contributions
(c.f. Fig.2a) from the remaining (``regular'') terms. We write
\begin{eqnarray}
A_i(\omega,\theta)&=&A_i(\omega,\theta)^{\pi^0-pole}+A_i(\omega,\theta)^{
                     regular} \\
 & & \quad i=1...6 \nonumber
\end{eqnarray}
The anomalous contributions to ${\cal O}(\epsilon^3)$ are given in Appendix
A for completeness. In the following we concentrate on the regular parts of
the amplitude.

One now performs a low-energy expansion\footnote{A Taylor expansion in
the energy for the anomalous parts of the amplitude is problematic due to
the rapid variateof the pion-pole contributions with energy. See Appendix A
for details.} of the six independent (regular)
structure
functions $A_i(\omega,\theta)^{reg.}$
in powers of the photon energy $\omega$. We
note that the $A_i$ are real functions for the case $\omega < m_\pi$, with
$m_\pi$ being the mass of the pion. For
the case of a proton target of mass $M_N$ with anomalous magnetic moment
$\kappa^{(p)}$ one finds
\begin{eqnarray}
A_1(\omega,\theta)_{c.m.}^{reg.}&=&-{e^2\over M_N}
-{e^2\over 4M_{N}^3}\left(1-\cos \theta\right) \omega^2 \nonumber \\
& & +4\pi \left( \alpha_{E}^{(p)}+
\cos \theta \; \beta_{M}^{(p)}\right) \omega^2 \nonumber \\
& & + \frac{4\pi}{M_N}\left(\alpha_{E}^{(p)}+\beta_{M}^{(p)}\right)
   \left(1+\cos \theta\right) \omega^3 \nonumber \\
& & + {\cal O}(\omega^4) \label{eq:a1} \\
A_2(\omega,\theta)_{c.m.}^{reg.}&=&{e^2\over M_{N}^2}\omega-4\pi\beta_{M}^{(p)}
\omega^2 \nonumber \\
& & - \frac{4\pi}{M_N}\left(\alpha_{E}^{(p)}+\beta_{M}^{(p)}\right) \omega^3
+{\cal O}(\omega^4) \label{eq:a2}\\
A_3(\omega,\theta)_{c.m.}^{reg.}&=&\left[1+2\kappa^{(p)}-(1+\kappa^{(p)})^2\cos
\theta\right]{e^2\over 2M_{N}^2}\omega\nonumber\\
& & + 4\pi \left[ \gamma_{1}^{(p)}-(\gamma_{2}^{(p)}+2\gamma_{4}^{(p)}) \cos
\theta \right]\omega^3\nonumber\\
& &-{(2\kappa^{(p)}+1)e^2\over 8M_{N}^4} \; \cos \theta \;
\omega^3+{\cal O}(\omega^4)
\label{eq:a3} \\
A_4(\omega,\theta)_{c.m.}^{reg.}&=&-{(1+\kappa^{(p)})^2e^2\over 2M_{N}^2}\omega+4\pi
\gamma_{2}^{(p)}\omega^3+{\cal O}(\omega^4)\\
A_5(\omega,\theta)_{c.m.}^{reg.}&=&{(1+\kappa^{(p)})^2e^2\over 2M_{N}^2}\omega+4\pi
\gamma_{4}^{(p)}\omega^3+{\cal O}(\omega^4)\\
A_6(\omega,\theta)_{c.m.}^{reg.}&=&-{(1+\kappa^{(p)})e^2\over 2M_{N}^2}\omega+4\pi
\gamma_{3}^{(p)}\omega^3 +{\cal O}(\omega^4) \label{eq:a6}
\end{eqnarray}
Note that for each structure function the leading order terms in the $\omega$
expansion are given by
{\em model-independent} Born contributions for scattering from a spin 1/2
point particle (with an allowed anomalous magnetic moment) and are
fixed by the low energy theorems (LET) of current algebra.  For example, in
the case of forward scattering
$\vec{k}=\vec{k}^\prime$ with the transversality condition
$\vec{\epsilon}\cdot \vec{k}=\vec{\epsilon}^{*\prime} \cdot \vec{k}=0$
only two of the six structure functions survive
\begin{eqnarray}
A_1(\omega,0)_{c.m}&=& 4 \pi \; f_1(\omega) \; ; \;\;\;\rightarrow f_1(0)_{\rm
LET}=-\frac{e^2 Z^2}{4\pi M_N} \\
A_3(\omega,0)_{c.m}&=& 4 \pi \omega \; f_2(\omega) \; ; \;\rightarrow
f_2(0)_{\rm LET}=-\frac{e^2\kappa^2}{8\pi M_{N}^2}
\end{eqnarray}
yielding the familiar Thomson result \cite{Th} with $Z=1$ for a proton
and the target spin-dependent LET
found many years ago by Gell-Mann, Goldberger and Low \cite{9}.

On the other hand, the higher order terms in $\omega$ are model-dependent
quantities
and the comparison here between theoretical predictions and
experimentally measured values provides an often sensitive
test of the validity of the
theoretical picture of the nucleon being employed.

Here, {\it e.g.},
the electric and magnetic polarizabilities $\alpha_E$ and $\beta_M$
discussed above enter the amplitude at ${\cal O}(\omega^2)$ and measure
the system's deformation in quasi-static
electric $(\vec{E})$ and magnetizing $(\vec{H})$ external fields \cite{10}
\begin{eqnarray}
\vec{d}=4\pi\alpha_E \vec{E} \; , \quad \vec{m} = 4\pi\beta_M \vec{H} \; ,
\end{eqnarray}
with $\vec{d}, \; (\vec{m})$ denoting the induced electric (magnetic)
dipole moment.

For the case of a microscopic target the situation gets more
complicated due to the possibility of target spin as an additional degree
of freedom that responds to external electric and magnetic fields. For
example, one can construct an induced spin-dependent dipole $\vec{p}_s$ via
$\vec{p}_{s} = \gamma_3 \vec{\nabla} ( \vec{S} \cdot \vec{B})$, where
$\vec{S}$ denotes the vector of the target spin and $\gamma_3$ corresponds
to a {\it spin-dependent polarizability} of the target. Ragusa has
analyzed\footnote{We have carried out the $\omega$-expansion of the structure
functions $A_i$ in the c.m. frame, whereas Ragusa's analysis was performed in
the Breit-frame. In the Breit frame the structure-dependent parts of
the amplitudes $A_1$ and $A_2$ are even functions of $\omega$.  When
transforming from the Breit frame to the c.m. frame we generate
additional terms which are odd in $\omega$.  However, one remains with
a total of two spin-independent and four spin-dependent quantities to
${\cal O}(\omega^3)$. For a detailed discussion we refer to \cite{structure}.}
the case of a spin 1/2 target \cite{Ragusa} and found that there exist four
independent quantities $\gamma_i,\quad i=1,\ldots 4$ which enter at
${\cal O}(\omega^3)$ and probe the spin-structure dependent pieces of the
Compton amplitude.  These ``spin-polarizabilities'' are perhaps less
familiar than their spin-independent counterparts $\alpha_E,\; \beta_M$
but they offer equally sensitive probes of nucleon structure.

Actually one combination of these terms {\it is} well-known.  If we
consider forward and spin-dependent scattering then one identifies the forward
spin-polarizability $\gamma_0$---
\begin{equation}
\gamma_0=\gamma_1-\gamma_2-2\gamma_4 \; . \label{eq:aa}
\end{equation}
The reason for the importance of this term is that if, based on Regge
arguments, one makes the assumption that the forward spin-flip
amplitude obeys an unsubtracted dispersion relation one finds
the result
\begin{equation}
f_2(\omega)={1\over 4\pi^2}\int_W^\infty
{ sds\over s^2-\omega^2}[\sigma_-(s)-\sigma_+(s)] \; , \label{eq:f2}
\end{equation}
where $\sigma_\pm$ are the photo-absorption cross sections for
parallel and anti-parallel alignments of the photon and target
helicities and $W=m_\pi + m_{\pi}^2/(2M_N)$ denotes the threshold energy for
an associated (neutral) pion in the intermediate state.  At $\omega=0$ this
becomes
\begin{equation}
{\pi e^2\kappa^2\over 2M_{N}^2}=-\int_W^\infty{ds\over s}[\sigma_-(s)-
\sigma_+(s)]
\end{equation}
which is the well-known Drell-Hearn-Gerasimov (DHG) sum rule and has
received a good deal of recent attention \cite{11}.

Differentiating Eq.(\ref{eq:f2}) with
respect to $\omega^2$ one finds a related sum rule for $\gamma_0$
\begin{equation}
\gamma_0={1\over 4\pi^2}\int_W^\infty{ds\over s^3}[\sigma_-(s)-\sigma_+(s)] \;
, \label{eq:g0}
\end{equation}
which was found originally by Gell-Mann, Goldberger and Thirring (GGT)
\cite{GGT}. The recent interest in the GGT sum rule Eq.(\ref{eq:g0}) has been
triggered by the fact that it relies on the same data sets required to test
the DHG sum rule thus providing a very sensitive check with even faster
high energy convergence due to the increased power in the energy denominator.

At present neither the forward spin-flip amplitude nor the spin-dependent
cross sections have yet been measured, but preparations are underway in
several laboratories around the world \cite{exp}. However, a model-dependent
analysis of the dispersive integral Eq.(\ref{eq:g0}) has been performed,
relying on multipole analyses of single-pion photoproduction data,
yielding\footnote{For claims that this multipole analysis is inconsistent
with the DHG sum rule see ref.\cite{SWK94}.} \cite{SWK94}
\begin{equation}
\gamma_0^{\rm disp.}\approx \left\{\begin{array}{ll}
-1.34\times 10^{-4}\,{\rm fm}^{4}& p \\
-0.38\times 10^{-4}\,{\rm fm}^{4}& n
\end{array}\right.\label{eq:dd}
\end{equation}
We shall return to this result below.

Finally, we want to note the recent determination
by the LEGS group of the corresponding spin-polarizability which one obtains
in the {\em backward} direction
\begin{equation}
\gamma_\pi=\gamma_1+\gamma_2+2\gamma_4 \; . \label{eq:gpi}
\end{equation}
They obtain \cite{6}
\begin{equation}
\delta_p \equiv - \gamma_\pi^{\rm exp}=(27.7\pm 2.3+2.8/-2.4)\times
10^{-4}\,{\rm fm}^4
\label{eq:bb}
\end{equation}
from a dispersion-based global fit to low energy Compton scattering
data,
yielding a dramatic difference in magnitude between the forward and
the backward spinpolarizabilities.
Now there exist some uncertainties in this determination because
higher order ({\it i.e.} $\omega^4$) structure dependent terms beyond
$\alpha_E, \;
\beta_M$ in the low-energy expansion of
$A_1(\omega,\theta)^{reg.},A_2(\omega,\theta)^{reg.}$ in
Eqs.(\ref{eq:a1},
\ref{eq:a2}) enter in the unpolarized cross section at the same order as the
spin-polarizabilities and are very poorly known at this point.
Nevertheless, we want to point out that the significant difference in magnitude
between $\gamma_{0}^{(p)}$ and
$\gamma_{\pi}^{(p)}$ can be easily explained once one connects these two
different linear
combinations of the four spin-polarizabilities with the underlying physics
using ChPT.

\section{ChPT Results}

The technical details of ${\cal O}(\epsilon^3)$ calculations in Compton
scattering with explicit nucleon and $\Delta$(1232) degrees of freedom
are discussed in ref.\cite{5}.  Here we give only the results,
in terms of coupling constants $g_A, \; b_1 \; ,g_{\pi N\Delta}$ defined
via the effective Lagrangians
\begin{eqnarray}
{\cal L}_{N}^{(1)}&=& \bar{N} \left( i v \cdot D + g_A S \cdot u \right) N \;
, \label{eq:l1} \\
{\cal L}_{N\Delta}^{(1)}&=&g_{\pi N\Delta}\bar{T}^\mu_iw^i_\mu N+ {\rm h.c.}
\; , \\
{\cal L}_{N\Delta}^{(2)}&=&{ib_1\over M_N}\bar{T}_i^\mu
S^\nu f_{+\mu\nu}^i N + \ldots \; , \\
{\cal L}_{\Delta}^{(1)}&=& - \bar{T}_{\mu}^i \left( i v \cdot D^{ij} - \Delta
\; \xi^{ij}_{3/2} + \dots \right) g^{\mu\nu} T_{\nu}^j \; , \\
{\cal L}_{\pi\pi}^{(2)}&=&\frac{F_{\pi}^2}{4}{\rm Tr} \left[ \left( \nabla_\mu
                          U \right)^\dagger \nabla^\mu U + \chi^\dagger U +
                          \chi U^\dagger \right] \; ,
                          \label{eq:l5}
\end{eqnarray}
where $i,j=1,2,3$ are isospin indices and $\xi^{ij}_{3/2}=\frac{2}{3}
\delta^{ij}-\frac{i}{3}\epsilon^{ijk}\tau^k$ is an isospin 3/2
projection operator with Pauli-matrix $\tau^k$. Here $U$ denotes a nonlinear
representation of the pion field with pion decay constant $F_\pi$ and
$N$ represents an (isodoublet) nucleon field of mass $M_N$. The delta degrees
of freedom are encoded in the fields $T^i_\lambda$ and are described in
terms of a Rarita-Schwinger representation both in spin and isospin space.
The mass-parameter
$\Delta$ can be chosen to correspond to the physical mass-difference between
nucleon and delta states and $S_\mu$ denotes the Pauli-Lubanski spin-vector
\begin{equation}
S^\mu={i\over 2}\gamma_5\sigma^{\mu\nu} v_\nu \; ,
\end{equation}
with heavy baryon four-velocity $v_\mu$.

The chiral field-tensors in Eqs.(\ref{eq:l1}-\ref{eq:l5}) are related to pion
$\pi^i$ and photon fields $A_\mu$ via
\begin{eqnarray}
D_\mu &=& \partial_\mu - i \frac{e}{2} (1+\tau_3) A_\mu + \ldots \; , \\
D_{\mu}^{ij}&=& \partial_\mu \delta^{ij} - i \frac{e}{2} (1+\tau^3) A_\mu
                \delta^{ij} + e \epsilon^{i3j} A_\mu + \ldots \; , \\
u_\mu &=& - \frac{1}{F_\pi} \tau^i \partial_\mu \pi^i + \frac{e}{F_\pi} A_\mu
          \epsilon^{i3j} \pi^i \tau^j + \ldots \; , \\
w_\mu^i&=&-{1\over F_\pi}\partial_\mu\pi^i-\frac{e}{F_\pi} A_\mu
          \epsilon^{i3j} \pi^j+\ldots \; , \\
f_{+\mu\nu}^i&=&e \; \delta^{i3} \left( \partial_\mu A_\nu - \partial_\nu
                A_\mu \right)+ \ldots \; , \\
\nabla_\mu U &=& \partial_\mu U - i \frac{e}{2} A_\mu \left[ \tau_3 , U \right]
                 + \ldots \; , \\
\chi &=& m_{\pi}^2 + \ldots \; ,
\end{eqnarray}
where $m_\pi$ is the mass of the pion in the limit of exact $SU(2)$ isospin
symmetry.

Furthermore, we need the anomalous $\pi^0\gamma\gamma$-vertex provided by the
Wess-Zumino-Witten Lagrangian \cite{WZW}
\begin{equation}
{\cal L}_{\pi^0\gamma\gamma}^{(WZW)}=-\frac{e^2}{32\pi^2F_\pi} \; \epsilon^{
\mu\nu\alpha\beta} \; F_{\mu\nu} F_{\alpha\beta} \pi^0 \; ,
\end{equation}
where $\epsilon_{0123}=1$ and $F_{\mu\nu}$ corresponds to the electromagnetic
field tensor.

Having defined the relevant lagrangians and coupling constants, we now present
the results of our ${\cal O}(\epsilon^3)$ calculation for the four
isoscalar spin-polarizabilities $\gamma_{i}^{(s)}$. All {\it regular}
isovector spinpolarizabilities $\gamma_{i}^{
(v)}$ are identically zero at ${\cal O}(\epsilon^3)$.
Choosing the heavy baryon velocity-vector
$v^{\mu}=(1,0,0,0)$ and working in the Coulomb gauge $v \cdot \epsilon = v
\cdot \epsilon^\prime =0$ we have calculated the contributions from $\pi-N$
continuum states (Fig.3), $\Delta$(1232) pole
graphs (Figs.2b+c) and $\pi-\Delta$ continuum states (Fig.4). We note that the
nucleon Born contributions up to ${\cal O}(\epsilon^3)$ (Fig.1) do not
contribute to the polarizabilities as defined in Eqs.(\ref{eq:a1}-\ref{eq:a6})
but provide the model-independent LETs of current algebra discussed in section
2. The (anomalous) contributions from neutral pion exchange (Fig.2a) are also excluded here and can be found in Appendix A.

For the $\pi -N$ continuum contributions (Fig.3) to ${\cal O}(\epsilon^3)$ in
the ``small scale expansion'' we find
\begin{eqnarray}
\gamma_1^{(s)\; N\pi}&=& + \; {e^2g_A^2\over 96\pi^3F_\pi^2m_\pi^2} ; \\
\gamma_2^{(s)\; N\pi}&=& + \; {e^2g_A^2\over 192\pi^3F_\pi^2m_\pi^2} ; \\
\gamma_3^{(s)\; N\pi}&=& + \; {e^2g_A^2\over 384\pi^3F_\pi^2m_\pi^2} ; \\
\gamma_4^{(s)\; N\pi}&=& - \; {e^2g_A^2\over 384\pi^3F_\pi^2m_\pi^2} ;
\end{eqnarray}
We note that our ${\cal O}(\epsilon^3)$ results for the $\pi -N$
continuum
agree with the ${\cal O}(p^3)$
HBChPT calculation of Bernard, Kaiser and Mei\ss ner \cite{7}. Finite shifts
to their $NN$
couplings due to the presence of explicit delta degrees of freedom will
only occur in higher order lagrangians and do not affect the $\pi NN$ coupling
$g_A$ to the order we are calculating. For a detailed discussion of this issue
we refer to ref.\cite{13}.

Next we give the ${\cal O}(\epsilon^3)$ contribution to
the spin-polarizabilities due to $\Delta$(1232) Born graphs (Figs.2b+c):
\begin{eqnarray}
\gamma_{1}^{(s)\; \Delta}&=& 0 ; \label{eq:pole1}\\
\gamma_{2}^{(s)\; \Delta}&=&
-{e^2\over 4\pi}{b_1^2\over 9M^2\Delta^2 } ;  \\
\gamma_{3}^{(s)\; \Delta}&=& 0 ; \\
\gamma_{4}^{(s)\; \Delta}&=&+{e^2\over 4\pi}{b_1^2\over 9M^2\Delta^2 } ;
\label{eq:pole4}
\end{eqnarray}
while for the ${\cal O}(\epsilon^3)$ $\pi -\Delta$ continuum terms (Fig.4)
we determine
\begin{eqnarray}
\gamma_{1}^{(s) \;\Delta\pi}&=&{e^2\over 4\pi}{g_{\pi N\Delta}^2\over
                               54\pi^2F_\pi^2}\left[-{\Delta^2+2m_\pi^2\over
                               (\Delta^2-m_\pi^2)^2}+{3\Delta m_\pi^2 \ln R
                               \over (\Delta^2-m_\pi^2)^{5\over 2}}\right]
                               ; \nonumber \\
& & \\
\gamma_{2}^{(s) \; \Delta\pi}&=&{e^2\over 4\pi}{g_{\pi N\Delta}^2\over
                                54\pi^2F_\pi^2}\left[{1\over \Delta^2-m_\pi^2}
                                -{\Delta\ln R \over (\Delta^2-m_\pi^2)^{3
                                \over 2}}\right] ; \nonumber \\
& & \\
\gamma_{3}^{(s) \; \Delta\pi}&=&{e^2\over 4\pi}{g_{\pi N\Delta}^2\over
                                108\pi^2F_\pi^2}\left[{1\over \Delta^2-m_\pi^2}
                                -{\Delta\ln R \over (\Delta^2-m_\pi^2)^{3
                                \over 2}}\right] ; \nonumber \\
& & \\
\gamma_{4}^{(s) \; \Delta\pi}&=&{e^2\over 4\pi}{g_{\pi N\Delta}^2\over
                                108\pi^2F_\pi^2}\left[-{1\over \Delta^2-
                                m_\pi^2}+{\Delta\ln R \over (\Delta^2-
                                m_\pi^2)^{3\over 2}} \right]  , \nonumber \\
\end{eqnarray}
with
\begin{equation}
R={\Delta\over m_\pi}+\sqrt{{\Delta^2\over m_\pi^2}-1} \; .
\end{equation}

For the parameter set $F_\pi=92.4$ MeV, $m_\pi=138$ MeV, $M_N=938$ MeV,
$\Delta=294$ MeV and the axial coupling $g_A=1.26$ determined from neutron
beta decay
we give the ${\cal O}(\epsilon^3)$ predictions for the spin-polarizabilities
in Table \ref{tab}. The $\pi
N\Delta$ ($\gamma N\Delta$) coupling constant $g_{\pi N\Delta}$ ($b_1$)
has been fixed from the experimental $\Delta$(1232) decay width and found to
be\footnote{It is important to note that these values are obtained by consistent use of 
the ``small scale expansion'' and are therefore appropriate for use
with other calculations at ${\cal O}(\epsilon^3)$.  In reference
\cite{5} considerably larger values $g_{\pi N\Delta}=1.5\pm 0.2, 
b_1^2=6.3\pm 1.75$, obtained from a tree level relativistic analysis, 
were employed.  We are now convinced that the correct procedure is the one
employed in the present work, so that the $\Delta$ effects obtained in
ref. \cite{5} should be appropriated rescaled downward by nearly a
factor of two.} \cite{13}
\begin{equation}
g_{\pi N\Delta}^{(1)}= 1.05 \pm 0.02 \; ;
\qquad b_{1}^{(2)}= 3.85 \pm 0.15 \; .
\end{equation}

As Table \ref{tab} clearly shows, $\gamma_{1}^{(s)},\gamma_{3}^{(s)}$ are
dominated by the contributions from the $\pi -N$ continuum (Fig.3)
in this ${\cal O}(\epsilon^3)$ calculation, whereas
$\gamma_{2}^{(s)},\gamma_{4}^{(s)}$ receive seizable corrections due to
delta pole graphs (Fig.2b+c).
At this order these are the only two of the four spin-polarizabilities
which receive contributions from delta pole exchange
(see Eqs.(\ref{eq:pole1}-\ref{eq:pole4})) via two successive magnetic
dipole (M1)
$\gamma N\Delta$ transitions. With respect to the spin-polarizabilities one
therefore has to perform at least an ${\cal O}(\epsilon^4)$ calculation in
order to be sensitive to the electric quadrupole (E2) $\gamma N\Delta$
transition moment. Furthermore, we note that the
effects of the $\pi - \Delta$ continuum (Fig.4) are found to be much smaller
than any of the other analyzed channels, as expected. It is interesting
to note that $\gamma_{2}^{(p,n)}$ displays a strong cancelation at this order
between
the $N\pi$-loop diagrams and the $\Delta$(1232) pole graphs, possibly
indicating a similar cancelation mechanism as the one postulated \cite{3,5}
for the magnetic polarizability $\beta_M$ at ${\cal O}(\epsilon^4)$! 

We now move on to discuss the connection with available experiments
pertaining to the spin-polarizabilities.

\section{Comparison with Experiment}

As mentioned above, so far there exists only limited experimental information
on two particular linear combinations of spin-polarizabilities $\gamma_i$.

In the forward direction we can compare the ${\cal O}(\epsilon^3)$
ChPT predictions with the multipole analysis of ref.\cite{SWK94}. Via
Eq.(\ref{eq:aa}) one finds
\begin{eqnarray}
\gamma_{0}^{(s) \; th.}&=& \left[ 4.6 \left( N\pi-{\rm loop}\right)
                           - 2.4 \left(
                           \Delta-{\rm pole} \right) \right. \nonumber \\
                       & & \left. \phantom{[ 4.6 ( N\pi}
                           - 0.2 \left( \Delta\pi-{\rm loop} \right) \right]
                           \times 10^{-4} \; {\rm fm}^4 \; , \nonumber \\
                       &=& + \; 2.0 \times 10^{-4} \; {\rm fm}^4 \; ,
\end{eqnarray}
where we have used the same parameter set as in the previous section.
We also note that all anomalous contributions from
Eqs.(\ref{eq:ae1}-\ref{eq:ae4}) in Appendix A cancel out exactly
to ${\cal O}(\epsilon^3)$
in this particular combination of spin-polarizabilities.
This makes the isoscalar combination $\gamma_{0}^{(s)}$ directly accessible
in experiments and thus provides a very sensitive test of the predicted
interference between
pion-nucleon and delta dynamics. At present, we can only conclude that in the
${\cal O}(\epsilon^3)$ calculation one obtains the forward spin-polarizability
of the proton with the opposite sign when comparing with the existing
multipole analysis Eq.(\ref{eq:dd}). Our delta pole contribution reduces the
large positive result of the $\pi -N$ continuum substantially, but at this
order we have no clear indication yet whether this trend will continue at
higher orders to lead to an overall negative result.

We note that the forward spin-polarizabilities have also been calculated
some time ago in relativistic one-loop ChPT, yielding \cite{BKKM92}
\begin{eqnarray}
\gamma_{0}^{(p) \; 1-{\rm loop}}&=& + 2.2 \times 10^{-4} \; {\rm fm}^4 \; ,
\\
\gamma_{0}^{(n) \; 1-{\rm loop}}&=& + 3.2 \times 10^{-4} \; {\rm fm}^4 \; .
\end{eqnarray}
However, we want to remind the reader that relativistic 1-loop ChPT does not
possess a systematic chiral power counting \cite{GSS} and only gives an
{\it indication} of some of the higher order corrections in the $\pi -N$
sector. In ref.\cite{BKKM92} it was further argued that the addition of
{\it phenomenological} delta exchange graphs can lead to negative results for
the
forward spin-polarizabilities in agreement with the multipole analysis
Eq.({\ref{eq:dd}). It remains to be seen whether a systematic calculation
to ${\cal O}(\epsilon^4)$ of all effects of $\Delta$(1232) contributions
will support this finding. On the experimental side a measurement scattering
circularly polarized photons off a polarized proton in the
forward direction would thus provide an independent check on the GGT sum rule
Eq.(\ref{eq:g0}) and the associated multipole analysis Eq.(\ref{eq:dd}).

Aside from this very limited information on the $\gamma_i$ in the forward
direction recently an analysis of the corresponding
spin-polarizability in the {\em backward
direction} has been reported \cite{6}.
Determining the pertinent linear combination
Eq.(\ref{eq:gpi}) of spin-polarizabilities accessible in this particular
angular direction we find the ${\cal O}(\epsilon^3)$ result
\begin{eqnarray}
\gamma_{\pi}^{(s) \; th.}&=& \left[ 4.6 \left( N\pi-{\rm loop}\right)
                           + 2.4 \left(
                           \Delta-{\rm pole} \right) \right. \nonumber \\
                       & & \left. \phantom{[ 4.6 ( N\pi}
                           - 0.2 \left( \Delta\pi-{\rm loop} \right) \right]
                           \times 10^{-4} \; {\rm fm}^4 \; , \nonumber \\
                       &=& + \; 6.8 \times 10^{-4} \; {\rm fm}^4 \; ,
\end{eqnarray}
in dramatic disagreement to the reported LEGS number Eq.(\ref{eq:bb}).
However, in addition to the uncertainty of higher order contributions affecting
the experimental extraction procedure noted in the previous section, we
believe the main origin of this discrepancy to arise from {\it anomalous}
contributions. While to ${\cal O}(\epsilon^3)$ these contributions cancel
exactly in the forward direction, they all interfere constructively and yield
a {\it maximal} contribution in the backward direction. From Appendix A one
finds for proton, neutron
\begin{eqnarray}
\gamma_{\pi}^{(p) \; anom.}&=& - \; 43.5  \times 10^{-4} \; {\rm fm}^4 \left( {\rm
                             anomaly} \right) , \\
\gamma_{\pi}^{(n) \; anom.}&=& + \; 43.5 \times 10^{-4} \; {\rm fm}^4 \left( {\rm
                             anomaly} \right)  ,
\end{eqnarray}
yielding
\begin{eqnarray}
\delta&=&- \left[\gamma_{\pi}^{(s)}+\gamma_{\pi}^{anom.}\right] \\
\rightarrow \delta_{p}^{th.}&=&+ \; 36.7 \times 10^{-4} \; {\rm fm}^4 , \nonumber \\
\rightarrow \delta_{n}^{th.}&=&- \; 50.3 \times 10^{-4} \; {\rm fm}^4 , \nonumber
\end{eqnarray}
which is roughly consistent with but is slightly larger in absolute
value than the reported backward ``spin-polarizability''
measurement on the proton Eq.(\ref{eq:bb}). Certainly a detailed analysis
of all higher order corrections in unpolarized backward scattering has to be
performed before one can draw strong conclusions on the discrepancy and
ultimately the result Eq.(\ref{eq:bb}) should of course
be checked in a polarized experiment. Nevertheless we have convincingly
shown that the
{\it regular} contribution of $\gamma_{\pi}$ to ${\cal O}(\epsilon^3)$
is of the same order of magnitude
as $\gamma_0$ and there is no theoretical reason to expect otherwise.
Any pion-nucleon and delta physics  will be obscured in a backward-measurement
unless it is possible to subtract the large anomalous contributions. If this
can be done, then it should be possible  to see the {\it constructive}
interference between the $\pi -N$ continuum contributions and the delta pole
graphs in $\gamma_{\pi}^{(s)}$, as opposed to the {\it destructive}
interference in $\gamma_{0}^{(s)}$ in the case of forward scattering!
However, this my require a fully polarized experiment.

It is certainly desirable to also perform polarized Compton scattering
experiments off polarized proton targets for angles $\theta$ different from
0 and 180 degrees. Analyses regarding the experimental feasibility are
in progress.\footnote{R. Miskimen and M. Pavan, private communication.}
These experiments would allow the determination of additional linear
combinations of the four spin-polarizabilities and finally lead to high
precision information on the complete set of spin-structure parameters of
the nucleon at low
energies. Theoretical investigations into favorable energy, angle and
spin-orientation
ranges are also underway and will be reported in a future communication.
However, in any case it is essential to extend the ChPT calculations
to ${\cal O}(\epsilon^4)$ in order to gain a better understanding of the
theoretical uncertainties.

\section{Conclusions}

We have in this note reviewed the theoretical status of the four nucleon
spin-polarizabilities $\gamma_i$ in the framework of ChPT as obtained in
polarized Compton scattering. We
have analyzed in detail the contributions of the $\pi -N$ continuum,
(anomalous)
neutral pion exchange, $\Delta$(1232) pole graphs and the $\pi-\Delta$
continuum. At this point connection with experimental information can only
be made for particular linear combinations of the spin-polarizabilities
accessible in forward or backward Compton scattering.
Our ${\cal O}(\epsilon^3)$ predictions are qualitatively consistent with the
new LEGS backward spin-polarizability number.
On the theoretical side an ${\cal O}(\epsilon^4)$
calculation is clearly called for to determine the convergence of the
perturbative series, whereas the upcoming Compton experiments with polarized
photons scattering off polarized nucleons should provide important information
on sign and magnitude of the individual spin-polarizabilities.

\acknowledgements

TRH would like to thank the theory group of the Institut
f\"{u}r Kernphysik for its hospitality and stimulating discussions
during a stay at Mainz where this work got started. Furthermore the
authors would like to acknowledge helpful discussions with Rory
Miskimen. This research
has been supported in part by the Natural Science and
Engineering Research Council of Canada, by the US National Science
Foundation, by Schweizerischer Nationalfonds and by Deutsche
Forschungsgemeinschaft SFB 201.

\appendix
\section{Anomalous Contributions}
The anomalous contributions to the six structure functions arising from the
Wess-Zumino-Witten functional \cite{WZW} read
\begin{eqnarray}
A_{1}^{\pi^0-pole}(\omega , t) &=& 0 \; ,\\
A_{2}^{\pi^0-pole}(\omega , t) &=& 0 \; , \\
A_{3}^{\pi^0-pole}(\omega , t) &=& + \; \frac{e^2 g_A}{8 \pi^2 F_{\pi}^{2}} \; \tau_3
                            \;  \frac{\omega t}{m_{\pi}^2 - t} \; , \\
A_{4}^{\pi^0-pole}(\omega , t) &=& 0 \; , \\
A_{5}^{\pi^0-pole}(\omega , t) &=& - \; \frac{e^2 g_A}{8 \pi^2 F_{\pi}^{2}} \; \tau_3
                            \; \frac{\omega^3}{m_{\pi}^2 - t} \; , \\
A_{6}^{\pi^0-pole}(\omega , t) &=& + \; \frac{e^2 g_A}{8 \pi^2 F_{\pi}^{2}} \; \tau_3
                            \; \frac{\omega^3}{m_{\pi}^2 - t} \; ,
\end{eqnarray}
with $t=-2 \omega^2 (1-\cos \theta)$ and $\tau_3$ being a Pauli matrix in
nucleon-isospin space.

Due to the small energy-denominators these functions are typically not
Taylor expanded in $\omega$ but kept in full as anomalous contributions.
If one wishes to extract contributions of this six amplitudes to the
spin-polarizabilities, one obtains
\begin{eqnarray}
\gamma_{1}^{(v)\;
{\rm anom}}&=&-{e^2g_A\over 16\pi^3F_\pi^2m_\pi^2}
            =- \; 21.7 \times 10^{-4} \; {\rm fm}^4, \label{eq:ae1} \\
\gamma_{2}^{(v)\;
{\rm anom}}&=&0 , \\
\gamma_{3}^{(v)\;
{\rm anom}}&=&+{e^2g_A\over 32\pi^3F_\pi^2m_\pi^2}
            =+ \; 10.9 \times 10^{-4} \; {\rm fm}^4, \\
\gamma_{4}^{(v)\;
{\rm anom}}&=&-{e^2g_A\over 32\pi^3F_\pi^2m_\pi^2}
            =- \; 10.9 \times 10^{-4} \; {\rm fm}^4, \label{eq:ae4}
\end{eqnarray}
which are all in the isovector channel at this order.

\begin{table}
\bigskip
\bigskip
\bigskip
\bigskip
\begin{tabular}{c|ccc||l}
$\gamma_{i}^{(s)}$ & $N\pi$-loop &$\Delta$-pole & $\Delta\pi$-loop & Sum \\
\hline
$\gamma_{1}^{(s)}$ & $+ 4.56$ & $0$ & $- 0.21$ & $+ 4.35$ \\
$\gamma_{2}^{(s)}$ & $+ 2.28$ & $- 2.40$ & $- 0.23$ & $- 0.35$ \\
$\gamma_{3}^{(s)}$ & $+ 1.14$ & $0$ & $- 0.12$ & $+ 1.02$ \\
$\gamma_{4}^{(s)}$ & $- 1.14$ & $+ 2.40$ & $+ 0.12$ & $+ 1.38$ \\
\end{tabular}
\bigskip
\caption{\label{tab} ${\cal O}(\epsilon^3)$ predictions for the isoscalar 
spin-polarizabilities $\gamma_{i}^{(s)}$. All results are given in the units 
$10^{-4}\;{\rm fm}^4$. All isovector spin-polarizabilities $\gamma_{i}^{(v)}$
are identically zero to this order.}
\end{table}

\end{document}